# Higher School of Economics
# (National Research University)

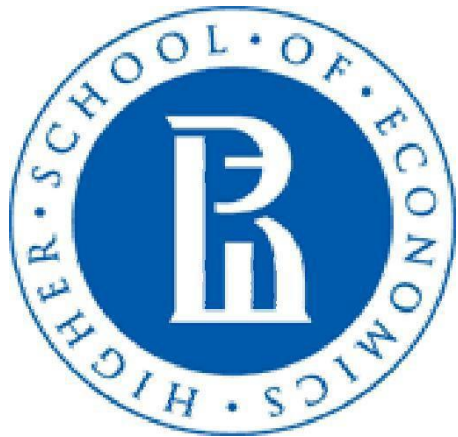

## Department of Software Engineering

## Course Work

## Topic:
## "Data Consistency Simulation Tool for NoSQL Database Systems"

**Supervisor:**

**Associate professor**

**Breyman Alexander**

**Davidovich**

**Presented by:**

**Nazim Faour**

# Contents

# ABSTRACT

Various data consistency levels have important part in the integrity of data and also effect performance especially the data that is replicated many times across or over the cluster. Based on BASE and the theorem of CAP tradeoffs, most systems of NoSQL have more relaxed consistency guarantees than another kind of databases which implement ACID. Most systems of NoSQL gave different methods to adjust required level of consistency to ensure the minimal numbering of the replicas accepted in each operation. Simulations are always depending on a simplified model and ignore many details and facts about the reality system. Therefore, a simulation can only work as an estimation or an explanation vehicle for observed behavior. So to create simulation tool, I have to characterize a model, identify influence factors and simply implement that depending on a (modeled) workload. In this paper, I gave a model of simulation to measure consistency of the data and to detect the data consistency violations in simulated network partition settings. So workloads are needed with set of users who make requests and then put the results for analysis.

# 1. INTRODUCTION

Recently, new epoch for high-performance database systems, low-cost, called NoSQL (Not Only SQL), has risen to challenge the strength of RDBMS. The principle components include: the possibility of scaling horizontally with ensuring low latency and high availability, data models and exible schemas, and simple low-level query interfaces instead of rich query languages. As a rule, the theorem of CAP and the model of PACELC explain the existence of direct trade-offs amongst consistency with availability or latency.

These trade-offs are a continuum, so there is currently a plenty of storage systems covering a wide scope of consistency guarantees. There are two different kinds of replication: the first called "Sync" replication which guarantee that all copies are new and the latest. It may be possible to have high latencies for the updates and can effect availability if the "Sync" replicated updates operations cannot be completed while another replicas are in offline status. The second type is Async replication which avoid the latencies of the high write but "Async" replication lets replicas to return a stale data. Practically speaking, many NoSQL systems relay on the high availability and the low latency and these systems apply a relaxed consistency strategy called eventual consistent which says that all of the replicas will, without any failures and updates operations, eventually reach a situation of consistent where all replicas are the same. In most of the situations where is no failure, the biggest size of inconsistency window can be bounded by many factors such as connection delays and how many replicas are required in the process of replication.

In real practice, the process of implementation and the eventually consistent strategy performance could be different between the systems depending on many factors such as replication of the data and synchronization protocols and the system load.

To create such an overall consistency benchmark, an evaluation of eventually consistent must be done. Such kind of that consistency benchmark must explain the relationship between consistency standards that are measured from both of the system perspective and the user perspective and the system performance within workloads and failures patterns. To guarantee these objectives, the database administration frameworks must Store the data constantly, Maintain data consistency and guarantee data availability.

It is important to guarantee that the correct data is written to a constant storing device. If the write or read operation does not precisely store or retrieve data, the database won't be of much use. This is seldom an issue unless there is a hardware failure. A more basic issue with reading and writing occurs when two or more users are using the database and want to do some operations on the same data at the same time. The consistency with as for database transactions refers to maintaining a single, logically cognizant perspective of data. Consistency has likewise been utilized to portray the condition of duplicates of information in appropriated framework Consistency also used to portray the condition of copies of data in distributed systems.

NoSQL Systems regularly execute eventual consistency; that is, there may be a timeframe where duplicates of data have diverse values, but eventually all duplicates will have the same value. This will lead to the possibility of a client to get different values from different servers in a cluster while querying to the database. NoSQL Systems usually use the concept of quorums when dealing with different operations such as reads and writes.

The number of servers that must respond to an operation (Write or read) until the operation will be considered complete called a quorum.

At the point When a read operation performed, the NoSQL system will read the data from numerous servers. Usually all of the servers will have consistent data. However, while the Database makes copies of the data from one of these servers to the others in order to store the replicas, the replica might have an inconsistent data.

There is an approach to see the right response for a read operation.

That way is to query all servers that storing the data. The database will check the number of distinct response values and then gives back the one that meets a configurable threshold. For instance, let's assume that the data in a NoSQL system is replicated to six servers and we have set the read threshold to 4. When four servers respond with the same reaction, the result of the query will be returned to the client.

We can vary the threshold to enhance the response time or the consistency. If the read threshold is set to 2, we will get such kind of fast response. We can say that the lower the threshold, the quicker the response but in other side the higher the risk of returning inconsistent data. Suppose that we decided to put the read threshold as 5, in this case we will guarantee consistent reading and the query will be returned only after all replicas have been updated. So it may lead to longer responding times.

Pretty much as we can conform a read threshold to balance response time and consistency, we can also modify a write threshold to balance response time and durability. Durability is the property of maintaining a right copy of data for long periods of time. A write operation is viewed as complete when a base number of replicas have been written to a constant storage.

If we put the write threshold as 2, then the writing operation will be finished when both servers write the data to a constant storage. This will lead to a quick responding times but in other hand poor durability. If that two servers leads to fail, the data will be lost. Let's say that we are working with the six-server cluster described previously. If data is replicated crosswise over four servers and we put the write threshold as 4, then all the four copies will be written to constant storage before the write operation finishes. If we set the threshold to 3, the data will be written to the three servers before the completing the write operation and the fourth copy will be written in a later time. If we decided to put the write threshold as at least 3 that will give the system the property of durability but if we put the number of replicas more than the threshold that will help to enhance the property of durability without increasing the response time of the writing operations.

This paper goal is to motivate and help the progress or the development of research on creating a standard model for quantifying consistency guarantees and the behavior of NoSQL systems.

In this paper, I attempted to characterize the fundamental requirements for building and designing this model and present the initial steps towards a complete consistency model.

Specifically, I summarize the fundamental contributions of this paper as follows:
 -The identification of the difficulties that should be considered in a complete consistency benchmark.
 -Analysis of state of the art consistency model of NoSQL systems.
 -The extension of a current model approach towards meeting the characterized consistency measurement challenges.

## 1.1 Related Work:

One proposing was by "Bailis et al" to simulate consistency in distributed systems and they introduce a model with definition as WARS [1].
Their own approach is bounded with the quorum systems by Dynamo style using the final write always win using the final Write always Win as a resolving inconsistency strategy [2]. However, my approach will allow for random replication schemes. To my understanding, their model suppose that same latencies distributions of the network link between all replicas and the model can just work for those replicas in single site. Moreover, they review neither ordering or client centric consistency manner nor failures.
An another work is dealing with inconsistencies of storing systems while using a component as a client side middleware and implemented this component as a library in order to increase the guarantees of an eventually consistent system [3]. For this reason, their work use Generalized Paxos with latency affects and under disadvantageous conditions for availability. On the top of Google BigTable [4] a Megastore [5] is implemented in order to increase the ensuring consistency via Paxos and 2PC. Both of the systems are selected for strong consistency guarantees which has higher influence on availability and performance than casual consistency guarantees, which I will propose in my approach, but also it may be alternative in the situation where more strong consistency guarantees are required. Another Work is for Bailis et al.'s [6] which maximize the staleness by reading just locally from the cache and bring high performance overheads.

## 2. Consistency Viewpoints:

To discuss the main point of consistency we should start by mentioning that there are two viewpoints of consistency in any distributed system [90]. The first is the provider, the entity that is responsible for the level of deployment and the operation process of a storing system, which will view the internal situation of the system. The main focus of a provider is on the synchronization processes between the replicas and how the operation are ordered. Hence, this viewpoint is called data-centric.

The other viewpoint is the client of the storing system. The client refers to a process that interacts with the storing system which can be any type of middleware, application or even any software that is running on the end user's machine. The client-centric focus is on the guarantees of the distributed storing system that can also be controlled as part of a contract between the provider and the end user that sets the level of the service that is expected from the provider.

Based on these two viewpoints, there are different consistency models either choosing a client-centric or choosing a data-centric. But there is a relation between those two models so that some of the models and combinations mean totally the same thing while still having different names.
Figure 1 demonstrates the different areas of interest for the data-centric and the client-centric consistency.

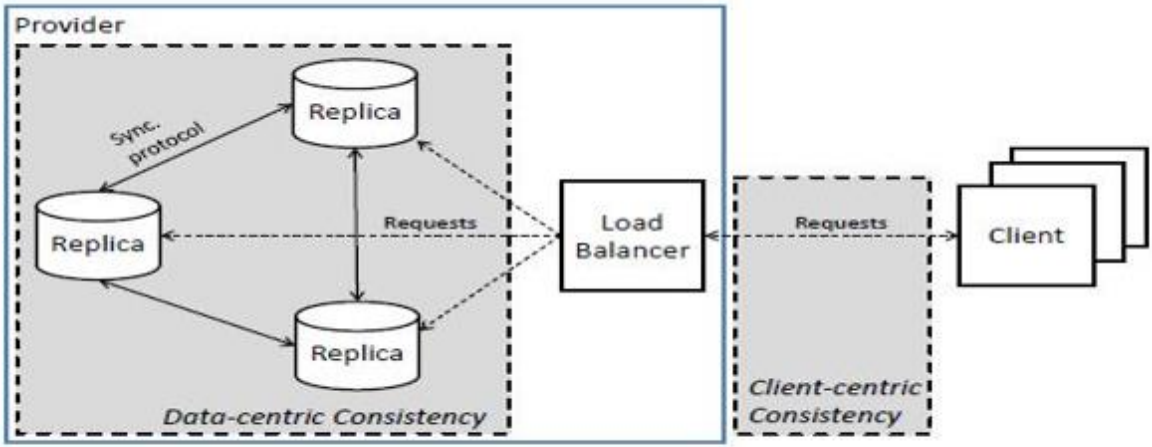

Figure 1 Data-centric and Client-centric Consistency

Both viewpoints have advantages and disadvantages for analyzing the consistency guarantees
relying upon the issue of interest. While the models of centric data consistency do not go for concrete implementations or algorithms, they obviously explain the ordering characteristic that permit to build up a corresponding synchronization protocol. The drawback is that models of centric data consistency are not so much helpful to the application developers.
Client-centric consistency models describe the main effects of that synchronization protocol. While this is very supportive to the application developer, it will completely ignore how that can be implemented.

## 3. Consistency Models:
Let's start by clarifying the models of centric client consistency before describing data-centric models and how those two are connected with each other. The client-centric models were suggested by Terry et al. [7].

## 3.1 Centric Client Consistency:

Monotonic Read Consistency (MRC) is the first model which ensures that if a client read a value n will from that point always read values >= n [8, 9]. In other words, MRC guarantees that if you make a query and then see the result, you will never see an earlier value of that query and when reading any new written value for the first time, all subsequent reads on this value will not return older values, figure 2.1.

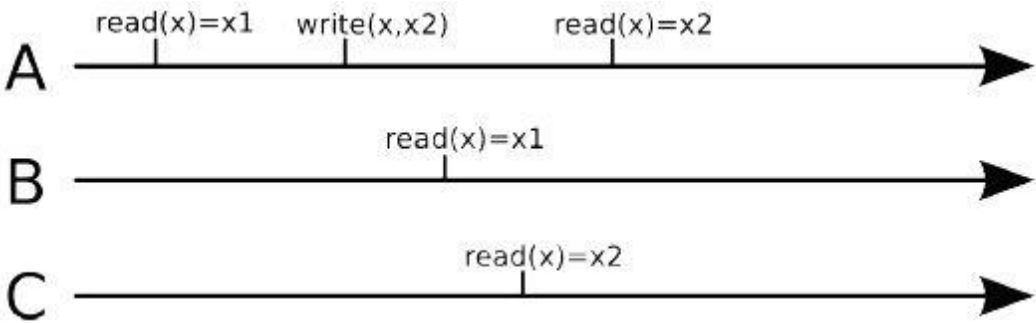

Figure 1.1 Monotonic read consistency: All processes will always see the latest version *of a data item that was read before from the database*.

This is useful as from an application point of view that data visibility might not be immediate but versions at least become visible in sequential order. So the system never "moves backward" in time.
For example, if a user A, who is using a system bank, sees the credited amount on his bank account statement and then he tries to exchange the money to a person B which fails due to "insufficient assets", this will at least cause severe client irritation if not more.

## Read your own writes(RYWC):

Read Your Writes Consistency (RYWC) ensures that a user that has written a version n will be able to read a version that is in any event as new as n [8, 9]. for instance, to keep away from client agitation when user A wants to check his own bank account statement, does not show the transaction and consequently wires the same amount of money again. RYWC will avoid any situations where a client or an application ordered same request many times because it will get idea that the request gone to failure situation in the first time. For idempotent operations reissuing requests causes just extra load on the system, while reissuing other requests will make serious inconsistencies.
In other words, once you have updated a record, all of your reads of that record will return the updated value. Figure 2.2 shows the behavior of a system that guarantees read-your-own-writes consistency.

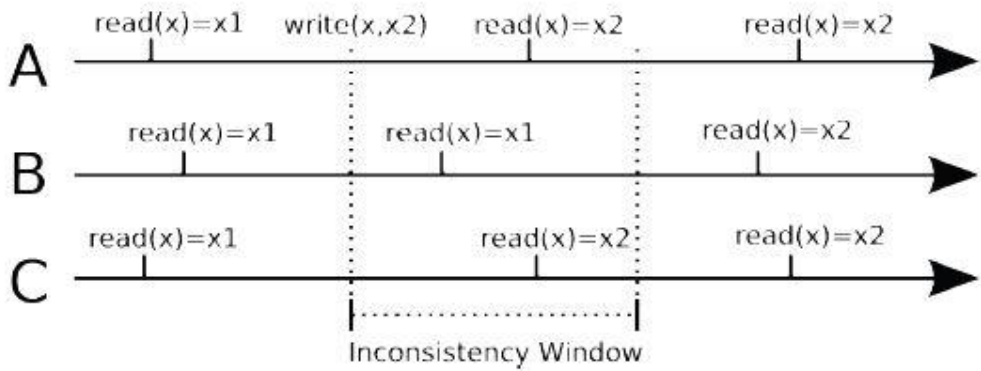

Figure2.2 Read-your-own-writes consistency: The writing process A will always reads its new value of the updated data item x, while other processes may *see some older value during the inconsistency window.*

**Monotonic Write Consistency (MWC)** ensures that any number of updates made by the user will be executed exactly in the same order that they have been requested [8, 9]. This will be very helpful in order to avoid apparently lost updates when first the client writes and then updates the data but this update operation executed before the first write and because of that it is overwritten, Figure 2.3. For example, in a bank system user A may have corrected the account number of a client B before finalizing the transfer. If the MWC is not ensured in the system, the money can end up in the wrong account. As indicated by Vogels [9] "any System that do not ensure this consistency level will be called a system that is hard to program".

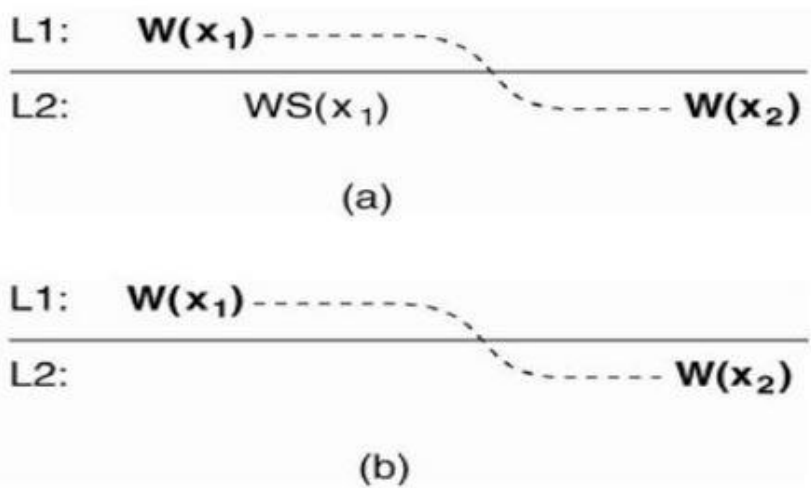

Figure 2.3 the write operations executed by a single process A at two different local copies of the same data store. (a) A monotonic-write consistent data store. (b) A data store that does not provide monotonic-write consistency.

**Write Follows Read Consistency (WFRC)** ensures that an update operation taking after a read operation of version n will be just executed on the replicas that are at any rate of version n [90]. on other words, if the client made a write operation on data item x following by a read operation on x by the same client then the write operation will be ensured to take place on the same or a more recent value of x that was read, Figure 2.4. This will be helpful against lost updates where the update is overwritten by a delayed update request for versions n. In NoSQL systems, these client-centric models are regularly not ensured explicitly but rather measurements show that they are satisfied for at least some parts of requests [10, 11].

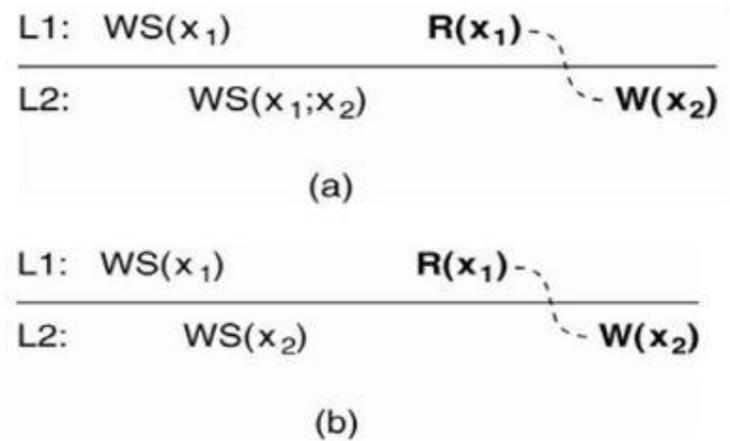

Figure 2.4 (a) A writes-follow-reads consistent data store. (b) A data store that does not provide writes-follow-reads consistency.

## 3.2 Data Centric Consistency:

There are Five Models:

### 1- Weak Consistency:

The guarantees are so weak and because of that they do not exist. Additionally, it means that replicas may by chance become consistent and it Does not give any ordering guarantees at all.

### 2- Eventual Consistency:

There can be inconsistent state in the NoSQL system in some period of time. This can happen when a client updates one copy of data and other copies continue to have the old version of the data. On other words, Eventual Consistency not ensure that each process will see same version of the value.

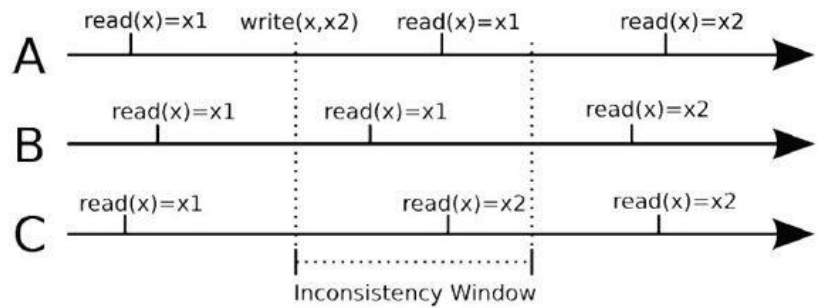

Figure 2.5 Eventual consistency: The processes A, B and C can see different versions of a data item during the inconsistency window, which is caused by asynchronous replication.

## Causal Consistency:

We can call it strong consistency because it is the most strict consistency Level that can be reached in a storing system [12]. In CC, all operations that contain causal demanding with another operation must be serialized in exact ordering on all replicas while unrelated operations may be serialized based on random order.

## Sequential Consistency:

Sequential Consistency (SC) is an extremely strict consistency model and it can't be accomplished in always accessible systems.

SC will require that all operations are serialized in the exact order on the whole replicas and this operations that is made by the same client will be executed in the same order that they were received by the storing system [8].

## Linearizability:

LIN explains what is the meaning of strict consistency that was mentiond before. LIN does not just consider ordering but also it consider the staleness. LIN demands that all of the non-concurrent requests are ordered by their entry time into the system and also all of the requests will always see the impacts of entirely preceding requests.

The figure 2.6 shows relationship between different client centric and data centric consistency models. 'N/A' means that the assurance can be reached for single requests from time to another but it will be based just on chance. 'Often' determines that such a behavior is seen for a big number of requests. 'Single Client' determines that the guarantees are satisfied, 'Global' is to explain when such a guarantee is reached out to all clients at the exact time.

| Data-centric Model | MRC | RYWC | MWC | WFRC |
|---|---|---|---|---|
| Weak Consistency | N/A | N/A | N/A | N/A |
| Eventual Consistency | Often | Often | Often | Often |
| Causal Consistency | Single Client | Single Client | Single Client | Single Client |
| Sequential Consistency | Single Client | Single Client | Single Client | Global |
| Linearizability | Global | Global | Global | Global |

*Figure 2.6 relationship Between Data-centric and Client-centric Consistency Models Ordered by the Strictness of their Guarantees*

## 3.3 Trade-offs in Consistency:

In any distributed storing systems, There are many trade-offs that exist and two of these trade-offs influence consistency in a direct way:

The theorem of CAP of Eric Brewer describes the harmony of consistency and availability, while The model of PACELC of Daniel Abadi extends the theorem of CAP for covering point of latency.

## 3.3.1 The CAP Theorem:

The CAP theorem, define three important features the Consistency, the Availability and tolerance to the network Partitions. This theorem has a view point that is impossible for any system to provide all three features in the exact time. The meaning of Availability is that the clients have ability to always write and read data in a particular time. The meaning of a partition tolerant between distributed systems is the failure tolerant versus problems of transitory connection and to allow the separation of the portions of nodes.

A system, that has the feature of partition tolerant, can only give strong consistency if the system has some reduction in its availability, in order to guarantee that every write process is accepted if the data has been copied to all needful nodes, but in distributed systems this approach of strong consistency is not always possible because of some connection problems and some transitory hardware failures. In distributed systems partitioning will happen most of the time so the only choice that will be left is between the consistency and the availability.

We can explain that by thinking up of a status with four replica and let's suppose that one of them is unable to be reached either because of some error in the server or some connection problems in the network. Let's say that an update operations has been requested in one of these replicas. The system in this situation has two choices: one of these choices is to execute the update only in three replicas and trade off consistency or the other choice is to dismiss the update with doing maintain for the consistency and delete the availability. in the same scenario if a client request read operation at one of the replicas, The system can respond with an error (like it is not able to read the unreachable replica which may contain a newer value) or the system can respond with just reading the reachable replicas. Figure3 gives a case of a replica on the right side that is unreachable and in the same time an update request in the other replica on the left. Here the system can choose the left way and refuse the request or go with the right way and sacrifice the consistency.

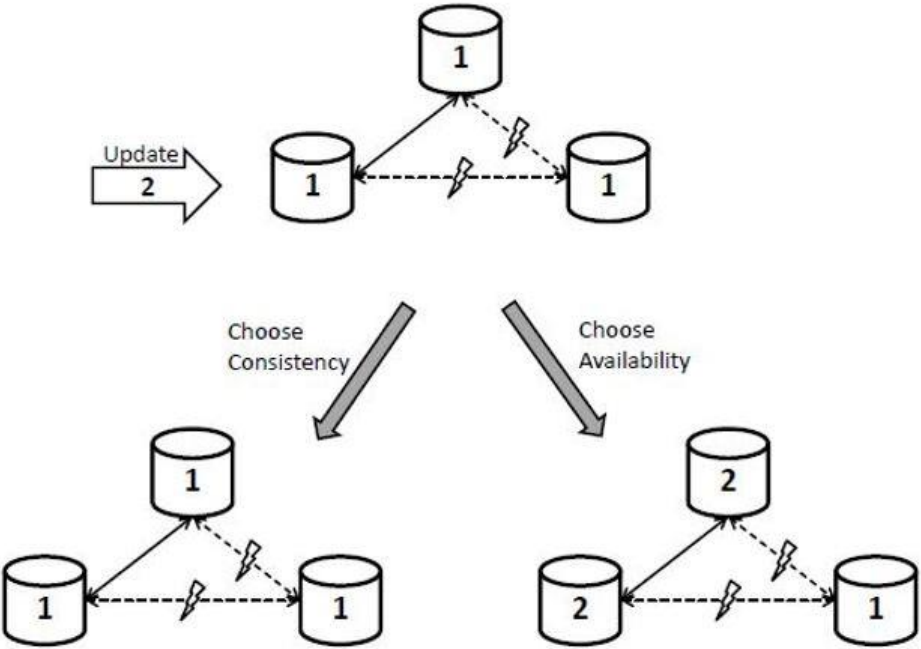

Figure 3 Consistency versus Availability Trade-off during Updates

## 3.3.2 The Model of PACELC

Daniel Abadi says that the theorem of CAP is far from the main case because CAP focuses on the tradeoff between consistency and availability and says that the main reason when NoSQL free consistency is to gain availability [13]. As a general rule, systems regularly free the consistency even if there is no any partitioning in the network in fact these systems do not look that they need to free the consistency. This happened because of the consistency with latency trade-off which may happen while writes and reads operations.

Let's take a scenario of four replicas that are distributed geographically, it will take some time if a client try to apply updates for all four replicas and wait until receiving acknowledgments. The designers of that system can update all four replicas in synchronous procedure and in that case they will be able to maintain the consistency but they will accept the high latencies in the system or the designers can updates all four replicas in the background by asynchronously procedure after the operation being committed already. For this situation the consistency will be sacrificed in order to gain low system latencies. Figure 3.1 is describing a case with two replicas while an operation of update: The system has two choices either go with update with accepting latency between the first and the second replicas or accepting the inconsistency window with low latency.

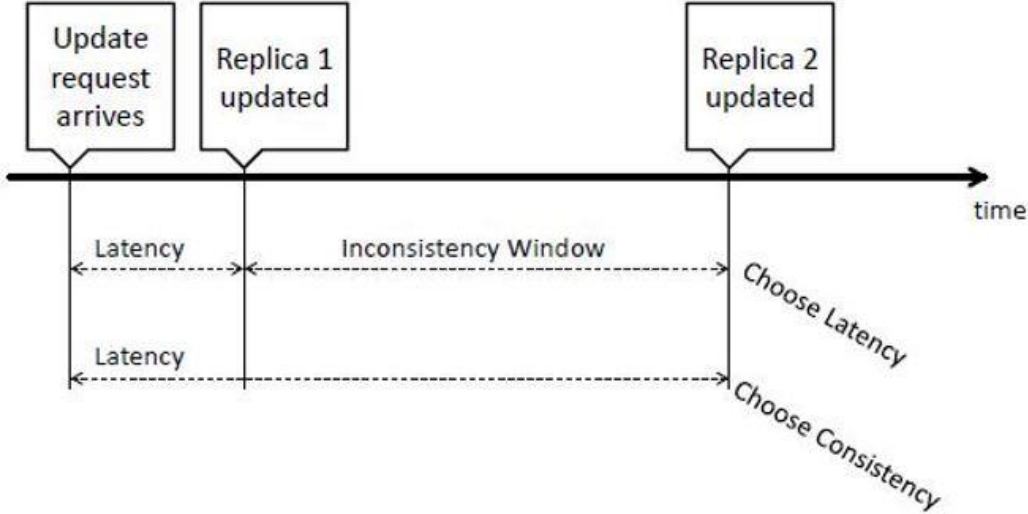

Figure 3.1 Consistency versus Latency Trade-off during Updates

This model (PACELC) is an extension of the theorem of CAP So If a partition exist in the system, the trade-off either can be between Consistency and Availability or between Consistency and Latency [13, 14].

### 3.3.3 BASE:

The BASE goal is to relax consistency for availability and performance. In other side, the applications should be designed in order to let them interact with inconsistencies results. For example, there are authoritative message queues that can be used either to emphasize that update operations reached the remotely replicas or in order to apply some level of ACID consistency while affecting more than one component of the data. The Systems, which solve suck kind of tradeoffs for the performance and availability, are usually mentioned as "BASE" which means Basically Available, Soft state and eventually consistent [15].

When a partial failure in some partitions of the distributed system appears but the rest part of the system continue to operate normally, this approach will be called basically available (BA). For instance, if a NoSQL systems is operating on 20 servers without any replicating data and one of these servers fails, then 20% of the clients' queries will lead to fail, but 80% will success. NoSQL systems usually create multiple copies of then data in different located servers. This BA will allow the system to respond for the requests and operations even if a failure appears in one server of them. When eventually more recent data appear after overwrite some old data, this approach will be called soft state (S). The soft state (s) property in BASE approach will interfere with the last property eventual consistency.

| ACID | BASE |
|---|---|
| Strong consistency | Weak consistency – stale data OK |
| Isolation | Availability first |
| Focus on "commit" | Best effort |
| Nested transactions | Approximate answers OK |
| Availability? | Aggressive (optimistic) |
| Conservative (pessimistic) | Simpler! |
| Difficult evolution (e.g. schema) | Faster |
| | Easier evolution |

*Figure* 2.2 ACID VS BASE

## 3.4 Data Consistency Level:

Write Consistency Levels:

| Level | Description | Usage |
|---|---|---|
| ONE | Returns a response from the closest replica, as determined by the snitch. By default, a read repair runs in the background to make the other replicas consistent. | Provides the highest availability of all the levels if you can tolerate a comparatively high probability of stale data being read. The replicas contacted for reads may not always have the most recent write. |
| TWO | Returns the most recent data from two of the closest replicas. | Similar to ONE. |
| THREE | Returns the most recent data from three of the closest replicas. | Similar to TWO. |
| LOCAL_ONE | Returns a response from the closest replica in the local data center. | Same usage as described in the table about write consistency levels. |
| SERIAL | Allows reading the current (and possibly uncommitted) state of data without proposing a new addition or update. If A SERIAL read finds an uncommitted transaction in progress, it will commit the transaction as part of the read. Similar to QUORUM. | To read the latest value of a column after a user has invoked a lightweight transaction to write to the column, use SERIAL. Cassandra then checks the inflight lightweight transaction for updates and, if found, returns the latest data. |
| LOCAL_SERIAL | Same as SERIAL, but confined to the data center. Similar to LOCAL_QUORUM. | Used to achieve linearizable consistency for lightweight transactions. |

| Level | Description | Usage |
|---|---|---|
|  | quorum of replica nodes in the same data Center as the coordinator node. Avoids latency of inter-data center communication. | such as NetworkTopologyStrategy, and a Properly configured snitch. Use to maintain consistency locally (within the single data center). Can be used with SimpleStrategy. |
| ONE | A write must be written to the commit log and memtable of at least one replica node. | Satisfies the needs of most users because consistency requirements are not stringent. |
| TWO | A write must be written to the commit log and memtable of at least two replica nodes. | Similar to ONE. |
| THREE | A write must be written to the commit log and memtable of at least three replica nodes. | Similar to TWO. |
| LOCAL_ONE | A write must be sent to, and successfully acknowledged by, at least one replica node in the local data center. | In a multiple data center clusters, a consistency level of ONE is often desirable, but cross-DC traffic is not. LOCAL_ONE accomplishes this. For security and quality reasons, you can use this consistency Level in an offline data center to prevent automatic connection to online nodes in other data centers if an offline node goes down. |
| ANY | A write must be written to at least one node. If all replica nodes for the given partition key are down, the write can still succeed after a hinted handoff has been written. If all replica nodes are down at write time, an ANY write is not readable until the replica nodes for That partition have recovered. | Provides low latency and a guarantee that a write never fails. Delivers the lowest consistency and highest availability. |

**Table: Read Consistency Levels**

| Level | Description | Usage |
|---|---|---|
| ALL | Returns the record after all replicas have responded. The read operation will fail if a replica does not respond. | Provides the highest consistency of all levels and the lowest availability of all levels. |
| EACH_QUORUM | Not supported for reads. | Not supported for reads. |
| QUORUM | Returns the record after a quorum Of replicas from all data centers has responded. | Used in either single or multiple data center clusters to maintain strong consistency across the cluster. Ensures strong consistency if you can tolerate some level of failure. |
| LOCAL_QUORUM | Returns the record after a quorum of replicas in the current data center as the coordinator node has reported. Avoids latency of inter-data center communication. | Used in multiple data center clusters with a rack-aware replica placement strategy (NetworkTopologyStrategy) and a properly configured snitch. Fails when using SimpleStrategy. |

| Level | Description | Usage |
|---|---|---|
| ONE | Returns a response from the closest replica, as determined by the snitch. By default, a read repair runs in the background to make the other replicas consistent. | Provides the highest availability of all the levels if you can tolerate a comparatively high probability of stale data being read. The replicas contacted for reads may not always have the most recent write. |
| TWO | Returns the most recent data from two of the closest replicas. | Similar to ONE. |
| THREE | Returns the most recent data from three of the closest replicas. | Similar to TWO. |
| LOCAL_ONE | Returns a response from the closest replica in the local data center. | Same usage as described in the table about write consistency levels. |
| SERIAL | Allows reading the current (and possibly uncommitted) state of data without proposing a new addition or update. If A SERIAL read finds an uncommitted transaction in progress, it will commit the transaction as part of the read. Similar to QUORUM. | To read the latest value of a column after a user has invoked a lightweight transaction to write to the column, use SERIAL. Cassandra then checks the inflight lightweight transaction for updates and, if found, returns the latest data. |
| LOCAL_SERIAL | Same as SERIAL, but confined to the data center. Similar to LOCAL_QUORUM. | Used to achieve linearizable consistency for lightweight transactions. |

## 3.5 MongoDB

It is document-oriented NoSQL system which manages storing the data in such kind of structure called BSON with dynamic schemas.
Automatic sharing is how MongoDB facilities the scalability with auto partitioning the data on many servers in order to support the increasing in data and the demand for reading and writing operations. MongoDB uses master-slave strategy So If the master node stopped or crashed, the strategy will select a slave and make it the new master. It is possible in MongoDB to send read requests to slave nodes instead of sending the requests to master.

Different read inclinations give verity guarantees levels of consistency and tradeoff
s, for instance if we read only from slaves, the master node can be relaxed for heavy writing workloads [16]. MongoDB's sharded architecture Figure 3.1.1.
The replication strategy in MongoDb is asynchronous, so the nodes which act like a slave may not give the last version of data but it is possible to determine the total number of nodes that should query the write operation before giving the value for the client successfully. This will allow to deal with tunable consistency [17].

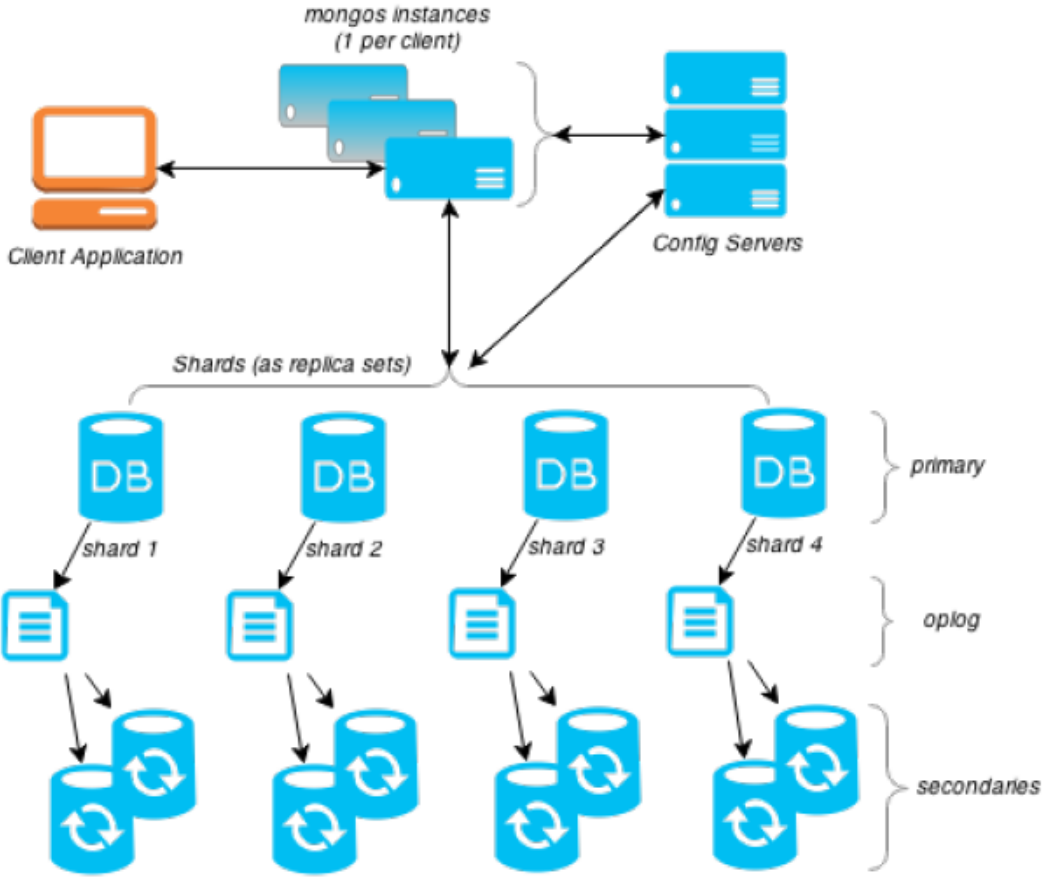

*Figure 3.1.1 MongoDb Archtiecture*

## 3.6 Cassandra

It is a Wide column storing system based on big table and dynamo DB.

The architecture of Cassandra is decentralized. It means that any operation can be executed by any node. From CAP theorem point it achieves two out of three, it achieves Availability and Partition Tolerance, Figure 3.1.2. It has very good performance of single row read. In order to reach strict consistency, it is required to apply quorum reads. Due to fact that it does not apply range based row scans, it will have some limitation in some specific situations. It will allow tunable consistency which give the chance to make tradeoffs amongst consistency and latency. For consistency levels the user can choose between different levels of consistency for every read and write request [18].

| Database | Category | Replication Strategy | Properties |
|---|---|---|---|
| MongoDB | Document | Replica-sets (Asynchronous) | CP |
| Cassandra | Extensible Record | Asynchronous Multi-Master | AP |

Figure 3.1.2 Mongo DB Vs Cassandra

## 3.7 Immediate Consistency:

One of the great aspects of Cassandra and MongoDB is that it has the ability to tune consistency among the difference between immediate consistency and EC (the eventual consistency). The great thing is you have a lot of control over so it's not just at a database level you can actually control immediate versus eventual on a query per query basis. So immediate consistency is a situation refers to a very specific situation. So In Cassandra immediate consistency is effectively when all of the replica nodes are guaranteed to have the same data at the same time. Immediate consistency can also be obtained by guaranteeing that a query will return the last version of the data. And this is really great so that even if your nodes your replica nodes are out of sync or slightly out of sync by controlling your read CL and you're write CL you can still gain what is effectively immediate consistency. Which basically means that every query will return the exact same and the most current data. Immediate consistency really refers to the ability of a select query to always return the last version of the data. Now that can either happen by controlling your CL levels you and by ensure that the data is the same on every node in the cluster that represent the replica. If the number of nodes written Plus the number of nodes read is greater than the replication factor Then you have immediate consistency. So if you have replication factor of three And you did a write CL as quorum Which means you're going to write to two nodes And if you're going to do read CL as quorum as well which means you're going to read from two nodes then you have a formula like if nodes written which is two plus nodes red which is two so it is four greater than replication factor three so you have four is greater than three then we have immediate consistency. Eventual consistency is a situation where you cannot guarantee that all replica nodes have the exact same data at the exact same point in time. It does not mean that all replica nodes do not have the same data it just means that When you tune for eventual consistency you no longer get a guarantee that the data is identical on all replica nodes And along with that is eventual consistency also means That any query is not guaranteed to return the most recent data now again. It is very often with eventual consistency all replica nodes will have the same data and very often the query will return the most recent data but again you're not getting a guarantee so if you have an SLA that you need to me for your application that requires the data is always fully consistent Then you would want to tune those queries for immediate consistency Where for other types of queries where performance is more important you can tune for eventual consistency. So again eventual consistency is effectively a scenario where select queries may or may not return the most recent data. When you go to deploy most Companies see eventual consistency numbers in a millisecond I mean sub second eventual consistency one of the best use cases or reported stats on this is actually from Netflix. Netflix reports there consistency in milliseconds so they use eventual consistency and the lag between when data is inconsistent to consistent is less than a second. So, when you need to obtain immediate consistency across your entire cluster, if you need all immediate consistency across all data centers across the entire cluster you have a few options. So the first one is you can set your write CL to all and you read CL to one. And in this scenario the reason you get immediate consistency is that by setting write cl to all you're insuring that all replica nodes have the most recent data so that you can query any one of them and still have immediate consistency. Now this configuration is read optimized you're going to get very fast reads But on the right side you're going to get some back pressure and some latency and you're also going to reduce your write throughput in exchange for increasing read throughput. Now next up you can do exactly the opposite and get immediate consistency and that is where you write CL to one and what that means is when you obviously the writes are going to fan out to all replica notes. But you're only going to get a response from one of them. And on the other side what you're going to do is when you read you're going to read from all now if you recall what's happening here is that the write may or may not be consistent in the three replica nodes however. Because you're read Cl is all. You're going to read from all three replica nodes the coordinators can emerge the data from all three replica notes and one of those three will be guaranteed to have the most recent Partition and therefore it'll merge those into what is effectively a single response with the most current and data the most current partition. This configuration is fully write optimized. So your writes are going to be faster you're going to have much better write throughput however the expense you're going to bear here is you're going to have a lower throughput and performance of your reads. Now of course if you recall you have full control over Write CL and read CL on every single request that you send and what this means is you can actually have one table that's read optimized in another one that's write optimized. And of course as you can imagine you have a third option, this is when you sat write CL to quorum, Read CL to quorum and you end up with a really balanced approach. When write CL to quorum it's going to write to a simple majority going write to two if all writes two of the three replica nodes. Reading from two of the three replica nodes. Now the nice thing about quorum quorum is you have High performance writes and high performance read and not getting the fastest performance of write of the fastest performance to read but you are getting high performance You're also getting good throughput on both again you're not going for the optimal solution either but you're getting good throughput. One important thing to mention here is you never want to do write CL all and read CL all it's completely unnecessary. If you're going to do write CL all do you read CL one and conversely if you do write CL one do you read CL all. And the reason you don't need all all is it's pretty much redundant you get no benefit from it and you're just going to have actually reduce throughput and performance as well as lower your availability it's basically just a horrible option that nobody should ever use. So, What about if you need immediate consistency in the local DC. What immediate consistency in the local DC means is that You're going to get a lot better performance with your application because what you're doing is if you have a cluster that has a data center in San Francisco another one in Chicago and another one in New York and you have users in New York, That users going to be accessing the New York data center. So by setting immediate consistency for the local DC. What you're doing is ensuring that users have full immediate consistency and their requests don't have to go to Chicago they don't have to go to San Francisco and so basically the fully optimizing for their local access now. The way you would achieve this if first as you sat write CL equals all and you said read CL equal to local one. Local one is just like one but it prevents the read request from going out to other data centers and stays within the local data center and again this is read optimized. you can do write CL local one where the writes will stay within the local data center the users in New York, Their writes would stay within the New York data center and the read request would go across the entire cluster. And this is going to be write optimize Which is really good when you have portions of your application that need very high write performance And of course the bounce is local quorum local quorum basically gives you all the benefit of quorum quorum Plus you're keeping your reeds and your writes within your local DC so you get immediate consistency with good tradeoff in performance now I'm not saying you should always use this because if you need to optimize for reads then do read CL local one if you need to optimize for writes then do write CL local one You have control of this on every single request you actually kind of application where parts of your some queries are read optimized other queries are write optimized and a third group Aquarius are balanced with high performance reads high performance writes. Beyond immediate consistency another very good option when you don't need immediate consistency is to use eventual consistency now. What this does is it gives you performance optimized. Eventual consistency when you set write CL one read CL one and that is eventual consistency across your entire cluster. And of course the best one you could choose is local one local one now if you recall your eventual is only a few milliseconds later so it's still less than a second and you're going to get really great performance you're going to get localized communication for your user as well as you're going to get the best read performance and the best write performance. You're also going to get the highest throughput and you're going to get the highest availability. So if you think about it by giving up a few milliseconds of consistency if your application can tolerate it with local one write CL and local one read CL you will get the best performance the best availability and the highest throughput.

| Write CL | Read CL | Description |
|----------|---------|-----------------|
| ALL      | ONE     | Read optimized  |
| ONE      | ALL     | Write optimized |
| QUORUM   | QUORUM  | Balanced        |

*Figure 3.2 three ways to achieve immediate consistency across the entire cluster*

| Write CL | Read CL | Description |
|---|---|---|
| ALL | LOCAL_ONE | Read optimized |
| LOCAL_ONE | ALL | Write optimized |
| LOCAL_QUORUM | LOCAL_QUORUM | Balanced |

*Figure 3.3 three ways to achieve immediate consistency on Local D*

## 4. Architecture of data consistency measurement tool

In order to measure the data consistency of NoSQL systems we need to build a framework. This framework architecture will be able to test how the parallel distributed workloads, distributed replication, the failures of nodes will influence the data consistency of a NoSQL system. For this reason, let's introduce the desired architecture with describing the components, Figure 4:

- ❖ Workload Builder:

This segment is utilized to build several workloads in order to allow the measurement of data consistency that get effected during the stages. It will try to report the results of latency or the throughput in order to define all tradeoffs between performance and consistency.

- ❖ data consistency Simulator(DCS):

The Simulator will be used to make a particular kind of simulation for the data consistency of a NoSQL system. While the Workload builder creates load, this part can build much more details about the behavior and the effects of the data consistency.

- ❖ Essential Measurement of Consistency (EMOC):

This part will be used to measure the behavior of data consistency. The output should use consistency from client viewpoint.

- ❖ Injection of Failure:

It will be used for causing some variation of failures.

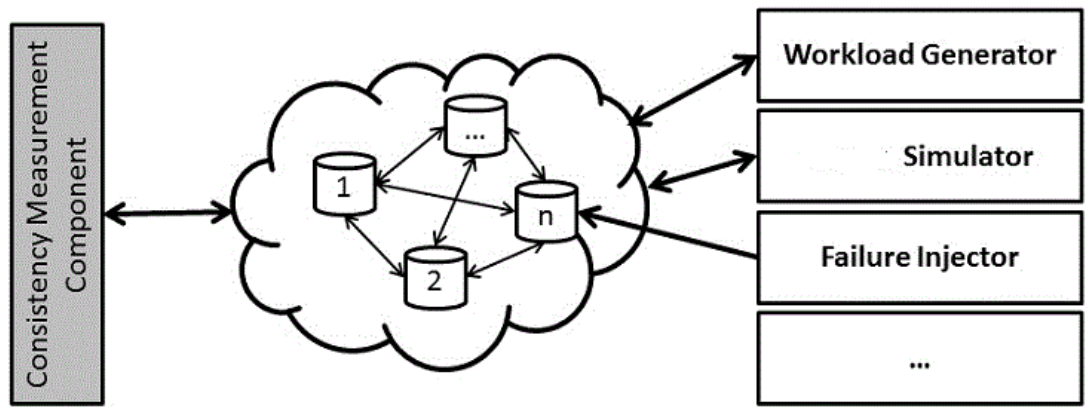

*Figure 4 Architecture of data consistency measurement tool'*

## 4.1 simulation tool:

To build a simulation tool we have to define the correspondence middleware that is used for the connections between the replicas.  We will also need number of servers in the whole different datacenters where we can deploy the replicas with any configurations. After that, we should deploy the simulation tool on the servers and finally run it. The tool also will collect data then terminate and download the whole results from servers.

In Distributed Systems, we also should try to estimate the servers failure averages and to measurement distributions of processing times. We will be able to model systems and configurations in simulation tool by using the information on the previous results, failures, and processing time. After that, it is possible to run the simulation tool and then analyze the output which contains the results.

## 4.2 The Model of Simulation Tool

The model will consist of three sub models, the basic model of the system which can be used separately of the other two models that are built on top of the basic model of the system, the model of failure and the cooperation model.

### 4.2.1 The basic model of the system

I can describe the basic model of the system by represent the connected replica or the connected servers by a graph $Gr=(R,C)$ where edges $C(Gr)$ represent the way of connections and vertices $R(Gr)$ represent the replicas or the servers. When we try to send the data over specific edge, it will take diverse period of time based on the bandwidth between any two vertices. The weights in edges will represent these time variations. $PD_{ij}(ds)$ represent probability of distribution for one path data transmit time(OPDTT), which explain the time required to send the data from node to another without any

"Ack" signal, while the sending process of the data with size of(ds) between vertex i to vertex j. it is not necessary that PDij(ds) will be always equal to PDji(ds).

### 4.2.2 The cooperation model

Gri=(R,C) represent directed subgraphs which explain the path of update operation for the specific replicas with knowing which replica has been updated and which one forwards the updates so it is possible to call that the graphs of replication. the same thing can be for the path of reading operation for knowing which the replica that has been read by another one and the order that represent the read request So it is possible to call that the graphs of reading. Each graph of replication/reading has value of probability which explain likelihood of read/write operation by using that specific graph (pWritei for the graphs of replication and pReadi for the graphs of reading). The values of probability will be effected by the way of the distribution of the replicas and the workload of the application and by the specific strategies of balancing the load. It is important to include in the cooperation model that the read/update operations can be Async, Sync, or element that consist in quorum.

Due to previous consideration, any edge of the graph of replication or the graph of reading will be related to one of three parts: synchronicity part, asynchronous part and quorum, some examples in these figures.

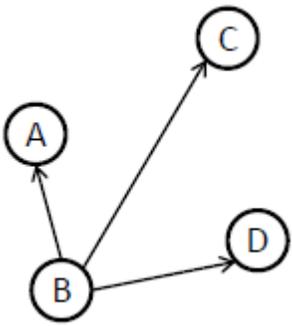 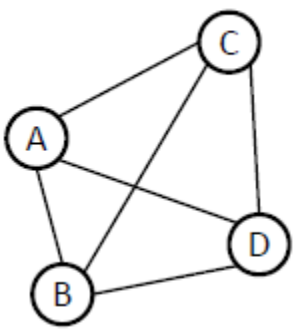 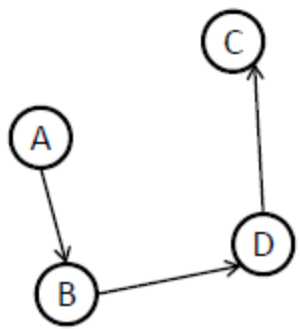

(a)The graph of Reading      (b) Connected replicas      (c) The graph of replication

### 4.2.3 The model of failure

We can define two types of failure for this model:
- The failure of stopping crash which describe that the failure happened because one component turns into unavailability with permanent situation.
- The failure of recovering crash describe that the failure happened because one component turns into unavailability within a specific and short period of time.

## 4.2.4 Simulation Strategies and Stages:

The simulation strategies are:

- The final write always win: all Updates will be procced strictly by the same order as these updates have been reached the particular replica But Cassandra applied different final write always win strategy where the users will timestamp the requests that they ordered and these timestamps are used to decide which one of the update requests was the Final request.
- Write set:
  The replica will be consistent on the off chance that it has all of the updates in an independent way of their order. Basically, writes will be attached to a log or being added to a set.
- the Returning of competing writes:
  There is a technique called vector clocks that system used it in order to
  The system uses mechanisms like vector clocks to distinguish the updates that are concurrent and then returns all conflict values based on a read. The following write operation, which is requested by the same client, will be expected to determine the conflict and solve it if no more writes operations happened in between [2].

In order to simulate the data consistency, I proposed three different stages:
- Stage 1:
  Creating big numbers of simulated requests. in every request, a graph of reading or a graph of replication will be chosen depending on the type of that request. For each request, a replication or read graph is chosen based on the request type (writing or reading) and the related values pWritei and pReadi. After that, graph traversal procedure will start in order to distinguish when the processing begins, when the processing ends and in which node. An event will be created for every start or end timestamp. Any delays because of some temporary failures will effect the Whole evaluation process of groups that are quorum synchronic. At the end, a log of the simulation will be created and it contains all events with the beginning and ending time separately for the reads and writes in vertices, the operation starting/ending or failing and a list of the vertices that are participating in a result of the graph of reading.
- Stage 2 :
  This stage will check the consistency of the data centric by first filtering all events that are in the log based on each event operation unique id and after that compute the variation in the time between both first and final timestamp. In this Stage, The output will be distribution of the data centric inconsistency windows, the error rate and latencies.
  So, the report of the Results will be per the graph of replication and the aggregation of all replication graphs (global).

- Stage 3 :
  This stage will check the consistency of the client centric. For this reason, an analyzing process have to be done for each simulated reading operation: depending on all strategies that have been mentioned above. The simulation tool will distinguish the writing operations that are contained in the response of that specific read operation or result of the operation based on the strategy. With doing a Comparison between this information and the request end timestamps of all the writes, which called commit timestamps, we will know if the read operation was stale. In addition, when setting this result in, association to all end timestamps of write requests that are done by the same user and all previous reads requests by same user, It can be easily decided if there were any violation in the MRC or the RYWC. This stage is important because it logs for every read request if there is a stale in the results were stale or there is a violation whether in MRC or RYWC. Also this stage will report for every write operation the request end timestamps of all the writes and the last time when the user seen read results which does not contain this certain write. Depending on that, a simple data spreadsheet analysis can define the violations probability of MRC and RYWC.

## 5. Conclusions and Future Work:

In my paper, I presented steps to create simulation model to check and measure data consistency guarantees of NoSQL Systems. I identified the requirements and the main challenges for that simulation model and gave architecture for a corresponding NoSQL system.

As a future work, this simulation model can be used to evaluate the effects of distributed replication servers among different geographical region and generate different kinds of workloads for two NoSQL systems Cassandra and MongoDB.

A set of data distributions like zipan and uniform can be applied in order to see the reality use cases with three levels of consistency such as QUORUM, ONE and ALL.  A Yahoo Cloud Serving tool can be used in future work as a workload builder in order to evaluate the data of different models of replication, of master-slave and multiple master models of replication. Amazon S3 can be used as cloud storing services with using a Cloud Watch to measure the Usage of the CPU for the replicas.